\documentclass{Interspeech}

\interspeechcameraready

\title{Facilitating Personalized TTS for Dysarthric Speakers Using \\ Knowledge Anchoring and Curriculum Learning}

\author[affiliation={1}]{Yejin}{Jeon}
\author[affiliation={1}]{Solee}{Im}
\author[affiliation={1}]{Youngjae}{Kim}
\author[affiliation={1,2}]{Gary Geunbae}{Lee}

\affiliation{GSAI}{POSTECH}{South Korea}
\affiliation{CSE}{POSTECH}{South Korea}
\email{jeonyj0612@postech.ac.kr, solee0022@postech.ac.kr, yj122198@postech.ac.kr, gblee@postech.ac.kr}

\usepackage{comment}
\usepackage{graphicx}
 \usepackage{amsmath}
\usepackage{subcaption}
\usepackage{multirow}
\usepackage{colortbl}
\usepackage{lipsum}  
\usepackage{bbding}
\usepackage{algpseudocode}
\usepackage{makecell}
\usepackage[ruled,vlined]{algorithm2e}
\newcommand\numberthis{\addtocounter{equation}{1}\tag{\theequation}}

\begin{document}

\maketitle

\begin{abstract}
    
    Dysarthric speakers experience substantial communication challenges due to impaired motor control of the speech apparatus, which leads to reduced speech intelligibility. This creates significant obstacles in dataset curation since actual recording of long, articulate sentences for the objective of training personalized TTS models becomes infeasible. Thus, the limited availability of audio data, in addition to the articulation errors that are present within the audio, complicates personalized speech synthesis for target dysarthric speaker adaptation. To address this, we frame the issue as a domain transfer task and introduce a knowledge anchoring framework that leverages a teacher-student model, enhanced by curriculum learning through audio augmentation. Experimental results show that the proposed zero-shot multi-speaker TTS model effectively generates synthetic speech with markedly reduced articulation errors and high speaker fidelity, while maintaining prosodic naturalness.
    
\end{abstract}

\section{Introduction}

Dysarthric speech emerges from a range of etiological factors, including cerebrovascular incidents such as strokes, as well as neuromuscular disorders linked to multiple sclerosis, cerebral palsy, and Parkinson’s disease \cite{Cause1, Cause2, Cause3}. These conditions compromise the neuromuscular control of speech production mechanisms, resulting in slurred, unintelligible, and phonetically distorted speech, which poses significant barriers to verbal communication. Given that speech is a fundamental medium for self-expression and social interaction, individuals with dysarthria frequently experience frustration, social isolation, and a diminished quality of life \cite{Life1}. 
In response to these communication challenges, research has taken two primary directions: improving comprehension from the listener’s perspective, and enhancing speech production from the speaker’s perspective. Most existing research has prioritized the former by conducting research focusing on the enhancement of automatic speech recognition (ASR) systems so as to facilitate more effective communication by enabling caregivers, medical professionals, and downstream assistive technologies to interpret dysarthric speech with greater accuracy \cite{ASR1, ASR2, ASR3, ASR4}.

While interpretation of affected speech is important, it is also crucial to empower dysarthric individuals by enhancing their ability to produce more intelligible speech and enhance their quality of life. Given that articulation deficits are a defining characteristic of dysarthria, early speaker-centric approaches sought to improve intelligibility by substituting impaired speech segments with unaffected ones \cite{Substitution1, Substitution2, Substitution3}. However, these approaches resulted in the loss of the speaker’s unique vocal identity as the average model voice was utilized. To mitigate this issue, voice banking emerged as an alternative, allowing individuals to pre-record speech samples before the onset of dysarthria for later reconstruction of their natural voice \cite{VoiceBanking1, VoiceBanking2}. Although effective for progressive degenerative conditions, this method is infeasible for individuals with sudden-onset dysarthria (e.g., stroke) as they lack pre-recorded samples.

Multi-speaker text-to-speech (TTS) synthesis offers an efficient alternative by generating speech from textual input while simultaneously cloning the speaker’s voice from reference audio. However, training such models from scratch or even employing few-shot learning using dysarthric speech presents significant challenges. Dysarthric speech often contains articulation errors, which if used as training data, result in models that produce unstable and unintelligible synthetic speech. Additionally, TTS models typically require large-scale, high-quality training datasets, which makes it impractical to rely solely on dysarthric recordings. To circumvent these issues, prior works have explored hybrid approaches that sequentially utilize a single-speaker TTS and voice conversion model \cite{Hybrid1, Hybrid2}.

\begin{figure}
\centering
\begin{subfigure}{0.37\textwidth}
    \includegraphics[width=\columnwidth, height=1.8cm]{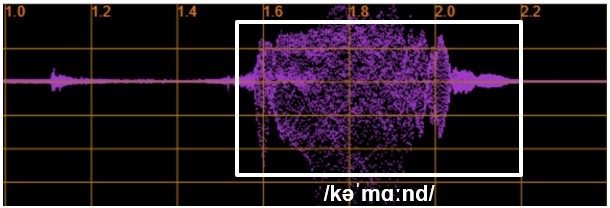}
    \caption{Normal articulation.}
    \label{fig:intro1}
\end{subfigure}
\hfill
\begin{subfigure}{0.37\textwidth}
    \includegraphics[width=\columnwidth, height=1.8cm]{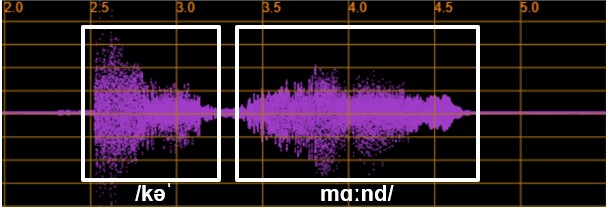}
    \caption{Dysarthric articulation.}
    \label{fig:intro2}
\end{subfigure}        
\caption{Visualizations of the articulation differences between a highly intelligible female speaker and a low-intelligibility dysarthric speaker. The dysarthric speaker exhibits elongated speech production (2.0s vs. 0.6s) and segmented articulation.}
\label{fig:Intro}
\end{figure}

In order to address the aforementioned limitations that are inherent in few-shot and from-scratch approaches, we focus on zero-shot multi-speaker TTS \cite{Zero1, Zero2} since just a single reference recording of the target speaker is required to conduct voice-preserving speech synthesis \cite{Zero1, Zero2}. Yet, extending this approach to the domain of dysarthric speech is not straightforward because there exists a fundamental mismatch between the audio samples used for model training and inference. This domain discrepancy is twofold; not only do variations in articulation exist, which range from highly intelligible to less intelligible speech, but dysarthric speech samples are often limited to single-word utterances \cite{DysData1} due to the speakers' physical difficulties in speech production (Figure \ref{fig:Intro}). This setting is in stark contrast to the much longer and well-articulated sentences typical of standard TTS datasets \cite{TTSData1, TTSData2}. Consequently, this challenge can be reframed as a zero-shot domain transfer problem, wherein the model must immediately and effectively extract speaker-specific vocal characteristics while remaining robust to articulation distortions in the input reference audio, which differ significantly from those encountered in the training data.

To resolve the dual domain transfer challenge seen in zero-shot multi-speaker TTS for dysarthric speakers, we take inspiration from pedagogical and physical therapy rehabilitation paradigms, where an expert plays a fundamental role in \textit{guiding} a learner in the acquisition of a skill. To the best of our knowledge, we propose the first teacher-student framework for this task, in which a teacher model generates an anchored representation that stabilizes learning and guides a student model throughout a structured learning process. Furthermore, a key aspect of this structured training is its adherence to a curriculum learning strategy, where the student model is gradually introduced to progressively shorter inputs. As a result, the student model learns to generalize effectively to the short and highly variable nature of dysarthric speech during inference. In essence, our methodology enables the model to disentangle speaker identity from speech articulation distortions, thereby facilitating the generation of highly intelligible speech that retains the distinct voice of the target dysarthric speaker. The efficacy of our approach is substantiated through both objective and subjective evaluations, and demonstrates its potential to enhance personalized communication for dysarthric individuals.

\section{Methodology}

\begin{figure}[t]
    \centering
    \includegraphics[width=0.99\columnwidth, height=5.3cm]{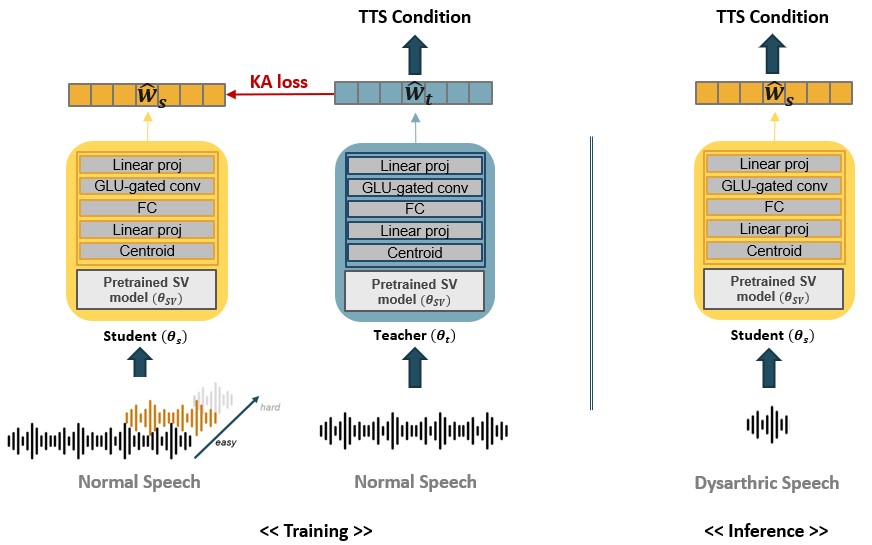}
    \caption{Training and inference processes. During training, the teacher model conditions the backbone TTS model while serving as an anchor for the student model. The student model is trained using curriculum learning.}
    \label{fig:arch}
\end{figure}

Our multi-speaker TTS model is composed of three main components: a module for generating a speaker representation from reference audio, a backbone text-to-mel-spectrogram acoustic model, and a vocoder that converts the mel-spectrogram into audio. The backbone acoustic model adheres to the FastSpeech2 architecture \cite{FastSpeech2}, which consists of a text encoder, a variance adaptor, and a decoder. To align with the objectives of multi-speaker TTS, we integrate the speaker representation that has been produced by the speaker encoder into the backbone TTS text encoder and decoder as in \cite{SALN}. Moreover, HiFi-GAN is employed as the pretrained vocoder \cite{HiFi-GAN} for mel-to-audio generation. Given that the ultimate goal is to achieve zero-shot extraction and utilization of speaker representations from dysarthric reference audio, we focus on the speaker encoder, which is structured around two key components: knowledge anchoring by a teacher for the student model, and curriculum learning specifically designed for the student model (Figure \ref{fig:arch}).

\subsection{Knowledge Anchoring}

Unlike conventional multi-speaker TTS systems, where the reference audio input during inference remains within the same domain as the training data, the task at hand requires effective extraction of timbre-specific features while remaining resilient to articulation distortions, as well as the scarcity of speaker information inherent in short audio segments at inference time. However, attempting to simultaneously address this dual-task objective using a single speaker encoder as in typical multi-speaker TTS systems \cite{TypSingEnc1, TypSingEnc2, TypSingEnc3}, is difficult. Therefore, to simplify this task, we separate it by introducing two collaborative teacher-student modules. A teacher speaker encoder initially captures the comprehensive information from the reference audio; this not only serves as an \textit{anchor} for the student module to learn from, but also as a preliminary \textit{filter} that removes irrelevant acoustic properties.
Both teacher $\theta_t$ and student $\theta_s$ models share an identical architecture, where the mel-spectrogram of the input audio $mel_{i,j}$ is passed through a speaker verification \cite{Resemblyzer} network\footnote{https://github.com/resemble-ai/Resemblyzer} $\theta_{SV}$ that is pretrained to compute the average embedding of each speaker. Formally, this is denoted as 

\begin{align*}
    e_{i,j} &= \theta_{SV}(mel_{i,j}) \numberthis \label{eqn1} \\ 
    c_i &= \frac{1}{M} \sum_{j=1}^{M} e_{i,j} \numberthis \label{eqn2}
\end{align*}

where \( c_i \) is the centroid for speaker \( i \), and \( e_{i,j} \) denotes the embedding of the \( j \)-th utterance from speaker \( i \). The resulting embedding $c \in \mathbb{R}^{256}$ is then used as the input to pass through a linear projection layer, a fully connected (FC) block with Mish activation \cite{Mish}, and a GLU-gated convolution stack with residual connections \cite{GatedCNN}. Another FC layer and temporal averaging results in two $N$-dimensional speaker style vectors $\hat w_{t} = \{y_1, y_2, \cdots, y_N\}$ and $\hat w_{s} = \{x_1, x_2, \cdots, x_N\}$ from each of the teacher and student modules, respectively. The following knowledge anchoring relationship is subsequently formed: 

\begin{equation}
    \mathcal{L}_{MAE}(\hat w_{s}, \hat w_{t}) = \frac{1}{N}\sum_{n=1}^{N}||x_n - y_n||_1
\end{equation}

\begin{algorithm}[t]
\label{alg:algorithm}
  \caption{Multi-view Augmentation for CL}
  \SetAlgoLined
  \SetKwInOut{Input}{Input}
  \SetKwInOut{Output}{Output}
  
  \Input{Audio $\mathcal{A}$, Total Training Steps $S$, Number of Cropping Stages $C$, Current Step $s$}
  \Output{Student Representation $\hat{w}_{s}$}

  \BlankLine
  \BlankLine
  \textbf{Step 1:} Compute Cropping Parameters \\
  \BlankLine
  Define steps of each cropping stage : $S_c = S / C$\;
  
  Compute cropping ratios for each stage:
  \BlankLine
$r = \begin{bmatrix} 1 - \frac{1}{C+1}, & 1 - \frac{2}{C+1}, & \dots, & 1 - \frac{C}{C+1} \end{bmatrix}$\;

  \BlankLine
  \BlankLine
  \textbf{Step 2:} Progressive Audio Cropping \\
  \BlankLine
  
    $\text{stage} \gets \lfloor \frac{s}{S_c} \rfloor$\;
    $\mathcal{A'} \gets \mathcal{A}[:r[\text{stage}] \times \text{len}(\mathcal{A})]$\;
    
  \BlankLine
  \Return $\hat{w}_{s} \gets \theta_s(\mathcal{A'})$ ;
  
\end{algorithm}

\subsection{Curriculum Learning (CL)}

Simply training the teacher and student models as described in the previous subsection presents two key challenges: (1) there is no direct connection between the training and inference domains, and (2) a trivial solution can emerge since both models share the same architecture and receive identical input data. To mitigate this, the student model is progressively trained with shorter segments of audio compared to the teacher model, enforcing a non-trivial learning process. Formally, let $S$ denote the total number of training steps and $C$ the number of progressive cropping stages applied to the input audio. The number of training steps allocated to each cropped audio stage is then given by $\frac{S}{C}$. An overview of this multi-view audio augmentation process is provided in Algorithm 1. 

Moreover, when training the entire TTS framework, the style representation $\hat{w}_{t}$ that is generated by the teacher model is employed as the condition for the backbone TTS encoder and decoder, while at inference time, the style representation generated by the student model $\hat{w}_{s}$ is utilized. Conditioning is implemented as in \cite{SALN}, where text input feature of the backbone text encoder $P_{text} = \{p_1, p_2, ..., p_K\}$ of K-dimensions is fused with style representation $w$ using its gain $g$ and bias $b$.

\begin{align*}
    \mu = \frac{1}{H}\sum_{k=1}^{H}p_k  &\text{ , } \sigma = \sqrt{\sum_{k=1}^{H}(p_k - \mu)} \numberthis \label{eqn5} \\ 
    N(p) &= \frac{p - \mu}{\sigma} \numberthis \label{eqn6}\\
    Fusion(P_{text}, \hat{w}) = &g(\hat{w}) \odot N(p) + b(\hat{w}) \numberthis \label{eqn7}
\end{align*}

The total loss is defined as a combination of the mel reconstruction loss between the synthesized audio and the original full reference audio used for the teacher model, and the loss between the style vectors that were computed by each of the teacher and student speaker encoders.

\begin{equation}
    \mathcal{L}_{Total} = \mathcal{L}_{MAE}(\hat {mel}, mel) + \mathcal{L}_{MAE}(\hat w_{s}, \hat w_{t}) 
    \label{eqn8}
\end{equation}

\section{Experimental Settings}

To validate our proposed methodology, we compare its performance against three zero-shot baseline models. The first employs discriminators trained with style prototypes and episodic training \cite{SALN}. While originally designed for normal speech, its method was also intended to generalize to unseen speakers, even those of short audio samples. On the other hand, the second and third baselines specifically target dysarthric speech: the former extracts a speaker representation using a pretrained speaker verification model \cite{Azizah}, while the latter adopts a hybrid approach akin to \cite{Hybrid1}. In this framework, a single-speaker FastPitch \cite{FastPitch} model generates speech corresponding to the target input text, and then re-synthesized to match the target dysarthric speaker’s voice using the FreeVC \cite{Freevc} voice conversion model.

To ensure fair comparisons, all experiments were conducted in identical conditions. Training was performed on a single GPU for the proposed 33M-parameter model using the Librispeech \cite{TTSData1} dataset with multi-view audio augmentation as described in Section 2.2. Specifically, audio cropping was conducted three times over 500,000 training steps (i.e., approximately every 160,000 steps). Zero-shot synthesis was conducted on the UASpeech dysarthria dataset \cite{DysData1} and evaluated by employing both subjective and objective metrics. MOS ratings were gathered from 19 Amazon Mechanical Turk workers who assessed naturalness in terms of prosodic intonation, and vocal similarity. Moreover, objective evaluations were conducted with phoneme error rate (PER), and speaker similarity, which was computed using cosine similarity between synthesized and ground truth speech representations as in prior research \cite{Sim-prior}.

\begin{table}[t]
\centering
\caption{Comparisons with Adaptive \cite{SALN}, Conditional \cite{Azizah}, and Hybrid \cite{FastPitch, Freevc} baseline models. 95\% confidence intervals are reported for MOS.}
\resizebox{0.92\columnwidth}{!}{%
\begin{tabular}{lcccc}
\midrule[1.2pt]
\multirow{2}{*}{\textbf{Model} \vspace{-0.5em}} & \multicolumn{2}{c}{\textbf{Obj. Metrics}} & \multicolumn{2}{c}{\textbf{Subj. Metrics}} \\ 
\cmidrule(r){2-3} \cmidrule(l){4-5}
& \textbf{PER ($\downarrow$)}  & \textbf{Spk Sim ($\uparrow$)} & \textbf{MOS-Nat} & \textbf{MOS-Spk} \\ 
\midrule
Adaptive  & 64.455   & 0.570  & 2.908 $\pm$ 0.23 & 2.753 $\pm$ 0.24  \\ 
Conditional & 47.696   & 0.647  & 2.839 $\pm$ 0.20 & 2.906 $\pm$ 0.21   \\ 
Hybrid  & 31.017   & 0.534  & 3.371 $\pm$ 0.15 & 3.731 $\pm$ 0.22  \\ 
\cellcolor[gray]{0.9}Proposed    & \cellcolor[gray]{0.9}14.254 & \cellcolor[gray]{0.9}0.619    & \cellcolor[gray]{0.9}3.601 $\pm$ 0.18 & \cellcolor[gray]{0.9}3.909 $\pm$ 0.21\\ 
\Xhline{1.2pt}
\end{tabular}%
}
\label{tab:baseline}
\end{table}

\begin{table}[t]
\centering
\caption{Subjective evaluations across different dysarthric speaker intelligibility (Int.) groups.}
\resizebox{0.92\columnwidth}{!}{%
\begin{tabular}{clcccc}
\midrule[1.2pt]
& \multicolumn{1}{c}{} & \multicolumn{1}{c}{\textbf{Very Low Int.}} & \multicolumn{1}{c}{\textbf{Low Int.}} & \multicolumn{1}{c}{\textbf{Middle Int.}} & \multicolumn{1}{c}{\textbf{High Int.}} \\ 
\cmidrule(lr){2-6}
\multirow{4}{*}{\rotatebox{90}{\textbf{MOS-Spk}}} 
& Adaptive & 2.722  & 3.053  & 3.000  & 3.000  \\  
& Conditional  & 2.789  & 2.947  & 2.895  & 2.763  \\  
& Hybrid  & 3.293  & 3.395  & 3.461  & 3.395  \\  
&\cellcolor[gray]{0.9}Proposed  &\cellcolor[gray]{0.9}3.338  &\cellcolor[gray]{0.9}3.763  &\cellcolor[gray]{0.9}3.618  &\cellcolor[gray]{0.9}3.882  \\  
\toprule
\multirow{4}{*}{\rotatebox{90}{\textbf{MOS-Nat}}}  
& Adaptive & 2.789  & 2.711  & 2.789  & 2.697  \\  
& Conditional  & 2.932  & 2.947  & 2.829  & 2.895  \\  
& Hybrid  & 3.699  & 3.737  & 3.763 & 3.750  \\  
& \cellcolor[gray]{0.9}Proposed  & \cellcolor[gray]{0.9}3.692  & \cellcolor[gray]{0.9}4.066  & \cellcolor[gray]{0.9}3.869  & \cellcolor[gray]{0.9}4.171  \\  
\toprule[1.2pt]
\end{tabular}%
}
\label{tab:BaseSpkGroup}
\end{table}

\section{Results and Discussion}
\subsection{Main Results}

The results presented in Table \ref{tab:baseline} underscore the efficacy of the proposed methodology in mitigating phonemic articulation errors while maintaining high speaker fidelity. PER demonstrates that the proposed methodology achieves a substantial reduction of over 50 points compared to the Adaptive \cite{SALN} baseline, and over 15 points compared to highest performing baseline model. Notably, this improvement of articulation accuracy does not come at the cost of speaker similarity, as the proposed model maintains a high speaker similarity score of 0.619, which is comparable to or surpasses existing approaches. Additionally, subjective evaluations reveal that the proposed method achieves the highest perceived naturalness in terms of intonation and pronunciation (MOS-Nat: 3.601) and speaker similarity (MOS-Spk: 3.909), which reinforces its effectiveness in preserving both articulation clarity and speaker identity.

\begin{table}[t]
\centering
\caption{Ablation studies on Knowledge Anchoring and Curriculum Learning. `Stu' denotes the student model, while `w/out CL' indicates direct training on the shortest audio inputs. CL is applied to the student model when present; otherwise, it is applied to the teacher model.}
\resizebox{0.92\columnwidth}{!}{%
\begin{tabular}{lcccc}
\midrule[1.2pt]
\multirow{2}{*}{\textbf{Model} \vspace{-0.5em}} & \multicolumn{2}{c}{\textbf{Teacher}} & \multicolumn{2}{c}{\textbf{Student}} \\ 
\cmidrule(r){2-3} \cmidrule(l){4-5}
& \textbf{PER ($\downarrow$)}  & \textbf{Spk Sim ($\uparrow$)} & \textbf{PER ($\downarrow$)} & \textbf{Spk Sim ($\uparrow$)} \\ 
\midrule
w/out Stu. w/out CL  & 26.428 & 0.618  & - & -   \\ 
w/out Stu. w/ CL      & 22.846 & 0.623 & - & -   \\ 
w/ Stu w/out CL      & 21.935   & 0.647  & 15.579 & 0.613   \\ 
\cellcolor[gray]{0.9}w/ Stu. w/ CL    & \cellcolor[gray]{0.9}20.559 & \cellcolor[gray]{0.9}0.646    & \cellcolor[gray]{0.9}14.254  & \cellcolor[gray]{0.9}0.619  \\ 
\Xhline{1.2pt}
\end{tabular}%
}
\label{tab:ablation}
\end{table}

\begin{table}[t]
\centering
\caption{Student-Teacher comparisons across different dysarthric speaker groups.}
\label{tab:Group-wise}
\resizebox{0.9\columnwidth}{!}{%
\begin{tabular}{lcccc}
\midrule[1.2pt]
\multirow{2}{*}{\textbf{Model} \vspace{-0.5em}} & \multicolumn{2}{c}{\textbf{Teacher}} & \multicolumn{2}{c}{\textbf{Student}} \\ 
\cmidrule(r){2-3} \cmidrule(l){4-5} 
& \textbf{PER ($\downarrow$)}  & \textbf{Spk Sim ($\uparrow$)} & \textbf{PER ($\downarrow$)} & \textbf{Spk Sim ($\uparrow$)} \\ 
\midrule
Very Low Int. & 20.528   & 0.586  & 15.157 & 0.556  \\ 
Low Int.      & 19.463   & 0.616  & 15.586 & 0.584  \\ 
Middle Int.   & 23.314   & 0.643  & 15.243 & 0.610  \\ 
High Int.    & 18.668   & 0.708  & 12.266 & 0.690  \\ 
\cellcolor[gray]{0.9}Average    & \cellcolor[gray]{0.9}20.559 & \cellcolor[gray]{0.9}0.646  & \cellcolor[gray]{0.9}14.254  & \cellcolor[gray]{0.9}0.619  \\ 
\Xhline{1.2pt}
\end{tabular}%
}
\end{table}

Beyond overall performance, Table \ref{tab:BaseSpkGroup} provides a breakdown of subjective evaluations across dysarthric speaker groups categorized by different intelligibility levels (Very Low, Low, Middle, and High). The proposed model consistently outperforms other approaches in speaker similarity across all intelligibility groups. In terms of naturalness, while the proposed model shows marginally lower scores than that of the Hybrid model for the Very Low category, the difference is only 0.007. Moreover, it can be seen that the proposed model surpasses the other baseline models for all remaining intelligibility levels.

\subsection{Knowledge Anchoring and CL}
The importance of the knowledge anchoring framework can be assessed by comparing models trained with and without the student model while maintaining curriculum learning in its most restrictive form (w/out CL in Table \ref{tab:ablation}). Note that this setting does not completely eliminate curriculum learning. Rather, it represents a scenario in which the speaker encoder is trained directly using the shortest audio from the multi-view augmentation process, and bypasses the gradual transition from longer to shorter references.
Under this condition, sole utilization of the teacher model as the TTS condition yields a high phoneme error rate (PER) of 26.428 (w/out Stu. w/out CL). However, introducing the student model (w/ Stu. w/out CL) leads to a substantial reduction in PER to 15.579, which marks an improvement of 10.849 points. Comparable speaker similarity scores are also retained between the teacher and student models.
A plausible explanation for these improvements is that the teacher model truly functions as a filter, which enables the student model to focus more effectively on speaker-specific timbre attributes. Thus, this reduces content leakage, since target timbre features do not inadvertently blend with miscellaneous acoustic factors, which would otherwise interfere with the target text's pronunciation when conditioning the backbone TTS model. These findings highlight the necessity of distinct teacher and student learning processes for enhancing phonemic articulation. 

While the teacher-student framework alone offers substantial benefits, its effectiveness is further amplified by curriculum learning. By structuring training such that the student speaker encoder initially learns from longer, more informative references before adapting to shorter, more constrained inputs, this gradual adaptation results in a PER that decreases from 26.428 (w/out Stu. w/out CL) to 22.846 (w/out Stu. w/ CL) for the teacher-only model, and from 15.579 (w/ Stu. w/out CL) to 14.254 (w/ Stu. w/ CL) for the knowledge-anchored model. Speaker similarity also increases from 0.613 to 0.619 for the student speaker encoder.
To further analyze model performance across dysarthric speaker groups, Table \ref{tab:Group-wise} presents results stratified by intelligibility level. A clear trend emerges; both teacher and student models achieve lower PER and higher speaker similarity as intelligibility increases. When using the teacher model to condition the backbone TTS, PER decreases from 20.528 (Very Low Intelligibility) to 18.668 (High Intelligibility), with speaker similarity improving from 0.586 to 0.708. Similarly, the student model’s PER drops from 15.157 to 12.266, with speaker similarity rising from 0.556 to 0.690. Notably, across all intelligibility levels, the student model consistently yields the lowest articulation errors. These results aligns well with the fundamental goal of speech synthesis, which is to foremost produce speech with accurate pronunciation. These trends are also intuitive, as higher intelligibility speakers exhibit fewer articulation distortions that obscure speaker-specific characteristics, which in turn leads to easier timbre extraction by the speaker encoder.

\begin{figure}[t]
    \centering
    \includegraphics[width=0.73\columnwidth, height=4.3cm]{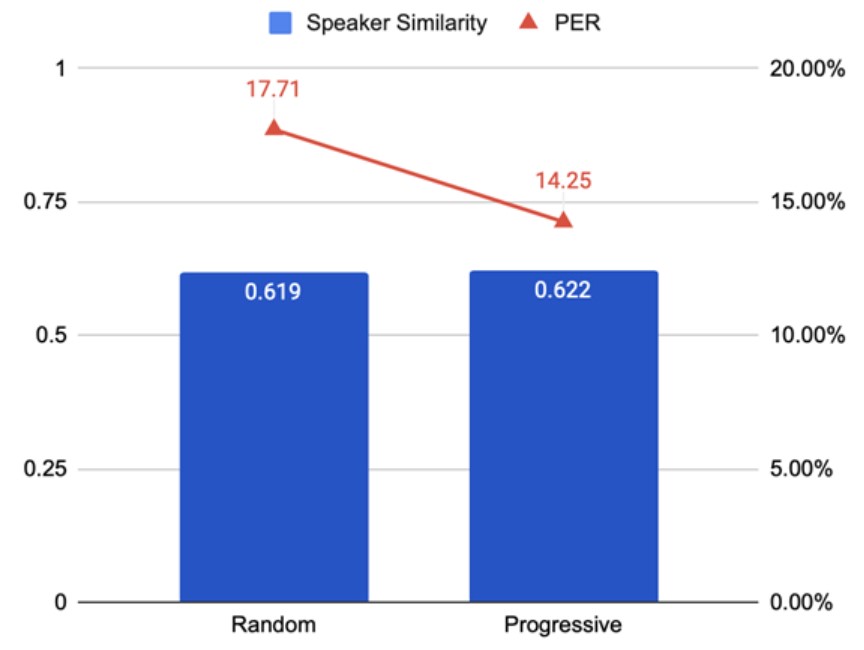}
    \caption{Variations of multi-view augmentation. Incrementally shortening audio significantly reduces PER while preserving speaker similarity comparable to the Random setting.}
    \label{fig:multiview.jpg}
\end{figure}

\subsection{Multi-view Audio Augmentation for CL}
In the proposed approach, the student model is trained using progressively shorter audio inputs to facilitate structured adaptation. To evaluate the efficacy of this augmentation strategy, we conduct an additional experiment using a randomized setting, where input audio utilized for the student speaker encoder is randomly cropped at each training step using one of the predefined cropping ratios. 
As shown in Figure \ref{fig:multiview.jpg}, the proposed progressive strategy leads to a substantial reduction in PER while preserving speaker similarity. Although the randomized setting exhibits a marginal increase in speaker similarity of 0.003, this is likely attributable to the greater proportion of longer audio inputs encountered during training compared to the structured progressive setting that was originally utilized.

\section{Conclusion}
In this paper, we present a zero-shot multi-speaker TTS approach to enhance personalized communication for dysarthric speakers. Unlike conventional single-encoder models, we introduce a knowledge anchoring framework with dual speaker encoders. Additionally, multiple views of the input reference audio is constructed and employed incrementally for curriculum learning. This approach effectively addresses the dual challenge of mitigating articulation artifacts while extracting robust speaker representations from minimal speech input. Comprehensive evaluations confirm that the synthesized speech remains highly intelligible, natural, and speaker-consistent.

\section{Acknowledgements}
This research was supported by Smart HealthCare for Police Officers Program(www.kipot.or.kr) through the Korea Institutes of Police Technology(KIPoT) funded by the Korean National Police Agency(KNPA, Korea)(No. RS-2022-PT000186; 47.5\%), by the IITP(Institute of Information \& Coummunications Technology Planning \& Evaluation)-ITRC(Information Technology Research Center) grant funded by the Korea government(Ministry of Science and ICT)(IITP-2025-RS-2024-00437866; 47.5\%), and by Institute of Information \& communications Technology Planning \& Evaluation (IITP) grant funded by the Korea government(MSIT) (No.RS-2019-II191906, Artificial Intelligence Graduate School Program(POSTECH); 5\%).

\bibliographystyle{IEEEtran}
\bibliography{template}

\begin{thebibliography}{10}
\providecommand{\url}[1]{#1}
\csname url@samestyle\endcsname
\providecommand{\newblock}{\relax}
\providecommand{\bibinfo}[2]{#2}
\providecommand{\BIBentrySTDinterwordspacing}{\spaceskip=0pt\relax}
\providecommand{\BIBentryALTinterwordstretchfactor}{4}
\providecommand{\BIBentryALTinterwordspacing}{\spaceskip=\fontdimen2\font plus
\BIBentryALTinterwordstretchfactor\fontdimen3\font minus \fontdimen4\font\relax}
\providecommand{\BIBforeignlanguage}[2]{{%
\expandafter\ifx\csname l@#1\endcsname\relax
\typeout{** WARNING: IEEEtran.bst: No hyphenation pattern has been}%
\typeout{** loaded for the language `#1'. Using the pattern for}%
\typeout{** the default language instead.}%
\else
\language=\csname l@#1\endcsname
\fi
#2}}
\providecommand{\BIBdecl}{\relax}
\BIBdecl

\bibitem{Cause1}
D.~Mulfari, G.~Meoni, M.~Marini, and L.~Fanucci, ``{Machine learning assistive application for users with speech disorders},'' in \emph{Applied Soft Computing}, 2021.

\bibitem{Cause2}
F.~Darley, A.~E. Aronson, and J.~R. Brown, ``{Differential diagnostic patterns of dysarthria},'' in \emph{Journal of Speech and Hearing Research}, 1969.

\bibitem{Cause3}
Y.~Yunusova, G.~Weismer, J.~R. Westbury, and M.~J. Lindstrom, ``{Articulatory movements during vowels in speakers with dysarthria and healthy controls},'' in \emph{Journal of Speech and Hearing Research}, 2021.

\bibitem{Life1}
J.~Mertl, E.~\u{Z}\'{a}\u{c}kov\'{a}, and B.~\u{R}epov\'{a}, ``{Quality of life of patients after total laryngectomy: the struggle against stigmatization and social exclusion using speech synthesis},'' in \emph{Disabil Rehabil Assist Technol.}, 2018.

\bibitem{ASR1}
S.~Leivaditi, T.~Matsushima, M.~Coler, S.~Nayak, and V.~Verkhodanova, ``{Fine-Tuning Strategies for Dutch Dysarthric Speech Recognition: Evaluating the Impact of Healthy, Disease-Specific, and Speaker-Specific Data},'' in \emph{Interspeech}, 2024.

\bibitem{ASR2}
I.-T. Hsieh and C.-H. Wu, ``{Dysarthric Speech Recognition Using Curriculum Learning and Articulatory Feature Embedding},'' in \emph{Interspeech}, 2024.

\bibitem{ASR3}
S.~Wang, S.~Zhao, J.~Zhou, A.~Kong, and Y.~Qin, ``{Enhancing Dysarthric Speech Recognition for Unseen Speakers via Prototype-Based Adaptation},'' in \emph{Interspeech}, 2024.

\bibitem{ASR4}
W.-Z. Leung, M.~Cross, A.~Ragni, and S.~Goetze, ``{Training Data Augmentation for Dysarthric Automatic Speech Recognition by Text-to-Dysarthric-Speech Synthesis},'' in \emph{Interspeech}, 2024.

\bibitem{Substitution1}
C.~Veaux, J.~Yamagishi, and S.~King, ``{Using HMM-based Speech Synthesis to Reconstruct the Voice of Individuals with Degenerative Speech Disorders},'' in \emph{Interspeech}, 2024.

\bibitem{Substitution2}
S.~Creer, S.~Cunningham, P.~Green, and J.~Yamagishi, ``{Building personalised synthetic voices for individuals with severe speech impairment},'' in \emph{Computer Speech \& Language}, 2013.

\bibitem{Substitution3}
Y.~Wang, X.~Wu, D.~Wang, L.~Meng, and H.~Meng, ``{UNIT-DSR: Dysarthric Speech Reconstruction System Using Speech Unit Normalization},'' in \emph{ICASSP}, 2024.

\bibitem{VoiceBanking1}
H.~T. Bunnell, J.~Lilley, C.~Pennington, B.~Moyers, and J.~Polikoff, ``{The ModelTalker System},'' in \emph{The Blizzard Challenge}, 2010.

\bibitem{VoiceBanking2}
J.~Yamagishi, C.~Veaux, S.~King, and S.~Renals, ``{Speech synthesis technologies for individuals with vocal disabilities: Voice banking and reconstruction},'' in \emph{Acoustical Science and Technology}, 2012.

\bibitem{Hybrid1}
K.~Matsubara, T.~Okamoto, R.~Takashima, T.~Takiguchi, T.~Toda, and Y.~Shiga, ``{High-Intelligibility Speech Synthesis for Dysarthric Speakers with LPCNet-Based TTS and CycleVAE-Based VC},'' in \emph{ICASSP}, 2021.

\bibitem{Hybrid2}
R.~Nanzaka and T.~Takiguchi, ``{Hybrid Text-to-Speech for Articulation Disorders with a Small Amount of Non-Parallel Data},'' in \emph{APSIPS Annual Summit and Conference}, 2018.

\bibitem{Zero1}
Y.~Jeon, Y.~Kim, and G.~G. Lee, ``{Enhancing Zero-Shot Multi-Speaker TTS with Negated Speaker Representations},'' in \emph{Proceedings of the AAAI Conference on Artificial Intelligence}, 2024.

\bibitem{Zero2}
B.~J. Choi, M.~Jeong, J.~Y. Lee, and N.~S. Kim, ``{SNAC: Speaker-Normalized Affine Coupling Layer in Flow-Based Architecture for Zero-Shot Multi-Speaker Text-to-Speech},'' in \emph{IEEE Signal Processing Letters}, 2023.

\bibitem{DysData1}
H.~Kim, M.~Hasegawa-Johnson, drienne Perlman, J.~Gunderson, T.~Huang, K.~Watkin, and S.~Frame, ``{Dysarthric Speech Database for Universal Access Research},'' in \emph{Interspeech}, 2008.

\bibitem{TTSData1}
H.~Zen, V.~Dang, R.~Clark, Y.~Zhang, R.~J. Weiss, Y.~Jia, Z.~Chen, , and Y.~Wum, ``{LibriTTS: A Corpus Derived from LibriSpeech for Text-to-Speech},'' in \emph{Interspeech}, 2019.

\bibitem{TTSData2}
K.~Ito and L.~Johnson, ``{The LJ Speech Dataset},'' in \emph{\url{https://keithito.com/LJ-Speech-Dataset/}}, 2017.

\bibitem{FastSpeech2}
Y.~Ren, Y.~Ruan, X.~Tan, T.~Qin, S.~Zhao, Z.~Zhao, and T.-Y. Liu, ``{FastSpeech: Fast, Robust and Controllable Text to Speech},'' in \emph{NeurIPS}, 2019.

\bibitem{SALN}
D.~Min, D.~B. Lee, E.~Yang, and S.~J. Hwang, ``Meta-stylespeech : Multi-speaker adaptive text-to-speech generation,'' in \emph{International Conference on Learning Representations (ICLR)}, 2021.

\bibitem{HiFi-GAN}
J.~Kong, J.~Kim, and J.~Bae, ``{HiFi-GAN: Generative Adversarial Networks for Efficient and High FIdelity Speech Synthesis},'' in \emph{NeurIPS}, 2020.

\bibitem{TypSingEnc1}
N.~Kumar and A.~N. andBrejesh Lall, ``{Zero-Shot Normalization Driven Multi-Speaker Text to Speech Synthesis},'' in \emph{IEEE/ACM Transactions on Audio, Speech, and Language Processing}, 2022.

\bibitem{TypSingEnc2}
E.~Casanova, J.~Weber, C.~Shulby, A.~C. Junior, E.~G\''{o}lge, and M.~A. Ponti, ``{YourTTS: Towards Zero-Shot Multi-Speaker TTS and Zero-Shot Voice Conversion for Everyone},'' in \emph{Proceedings of Machine Learning Research}, 2022.

\bibitem{TypSingEnc3}
E.~Cooper, C.-I. Lai, Y.~Yasuda, F.~Fang, X.~Wang, and N.~Chen, ``{Zero-Shot Multi-Speaker Text-To-Speech with State-Of-The-Art Neural Speaker Embeddings},'' in \emph{ICASSP}, 2020.

\bibitem{Resemblyzer}
L.~Wan, Q.~Wang, A.~Papir, and I.~L. Moreno, ``{Generalized End-to-End Loss for Speaker Verification},'' in \emph{ICASSP}, 2018.

\bibitem{Mish}
D.~Misra, ``{Mish: A Self Regularized Non-Monotonic Activation Function},'' in \emph{BMVC}, 2020.

\bibitem{GatedCNN}
Y.~N. Dauphin, A.~Fan, M.~Auli, and D.~Grangier, ``{Language Modeling with Gated Convolutional Networks},'' in \emph{Proceedings of the 34th International Conference on Machine Learning, ICML}, 2017.

\bibitem{Azizah}
K.~Azizah, ``{Zero-Shot Voice Cloning Text-to-Speech for Dysphonia Disorder Speakers},'' \emph{IEEE Access}, vol.~12, pp. 63\,528--63\,547, 2024.

\bibitem{FastPitch}
A.~{\L}ańcucki, ``{FastPitch: Parallel Text-to-speech with Pitch Prediction},'' in \emph{ICASSP}, 2021.

\bibitem{Freevc}
J.~Lii, W.~Tu, and L.~Xiao, ``{Freevc: Towards High-Quality Text-Free One-Shot Voice Conversion},'' in \emph{ICASSP}, 2023.

\bibitem{Sim-prior}
Y.~Zhou1, C.~Song, X.~Li, L.~Zhang, Z.~Wu, Y.~Bian, D.~Su, and H.~Meng, ``{Content-Dependent Fine-Grained Speaker Embedding for Zero-Shot Speaker Adaptation in Text-to-Speech Synthesis},'' in \emph{Interspeech}, 2022.

\end{thebibliography}

\end{document}